\documentclass[aps,prd,reprint,superscriptaddress,nofootinbib,floatfix]{revtex4-2}

\usepackage{amsmath}
\usepackage{amssymb}
\usepackage{booktabs}
\usepackage{graphicx}
\usepackage{xcolor}

\makeatletter
\let\pre@bibdata\@empty
\def\@bibdataout@init{}
\makeatother

\begin{document}

\title{A Boundary-Consistent Two-Zone Electron Kernel for Distant Pulsar Contributions to Positron Flux and Anisotropy}

\author{Yiwei Bao}
\affiliation{Tsung-Dao Lee Institute, Shanghai Jiao Tong University, Shanghai 201210, China}
\affiliation{School of Physics and Astronomy, Shanghai Jiao Tong University, Shanghai 200240, China}

\author{Jie-Shuang Wang}
\email{jieshuangwang@sjtu.edu.cn}
\affiliation{Tsung-Dao Lee Institute, Shanghai Jiao Tong University, Shanghai 201210, China}
\affiliation{School of Physics and Astronomy, Shanghai Jiao Tong University, Shanghai 200240, China}

\author{Hao Zhou}
\email{hao\_zhou@sjtu.edu.cn}
\affiliation{Tsung-Dao Lee Institute, Shanghai Jiao Tong University, Shanghai 201210, China}
\affiliation{School of Physics and Astronomy, Shanghai Jiao Tong University, Shanghai 200240, China}

\date{\today}

\begin{abstract}
We present a semi-analytical series solution for electron and positron
propagation in a spherical two-zone diffusion model.  The solution treats slow
diffusion inside a near-source region and standard interstellar diffusion
outside it, while synchrotron and Klein--Nishina inverse-Compton cooling are
included through energy characteristics.  The formulation avoids the
oscillatory cancellations of direct two-zone integral evaluations and preserves
the sharp radiative cooling boundary seen in finite-volume checks.

We apply the kernel to pulsar contributions to the local cosmic-ray lepton
flux.  Nearby pulsars remain natural candidates near the TeV cutoff, but at
tens to hundreds of GeV the larger source volume allows more distant pulsars to
contribute collectively: for a disk half-thickness of $0.2\,{\rm kpc}$, sources
beyond $1\,{\rm kpc}$ can still provide $37$--$47\%$ of the
$10$--$100\,{\rm GeV}$ flux.  Comparing with AMS-02 positron data and
all-electron anisotropy limits, and imposing an inner $100\,{\rm pc}$ cavity
motivated by the Local Bubble and pulsar proper motions, we find that
Geminga-scale slow-diffusion halos remain compatible with current data.  The
fitted pulsar component is dominated by sources beyond $0.3\,{\rm kpc}$, but
flux and anisotropy data alone do not uniquely determine the halo size;
external information such as TeV halo morphology is still required.
\end{abstract}

\maketitle

\section{Introduction}
\label{sec:introduction}

The local cosmic-ray electron and positron spectra have become precision probes
of nearby high-energy accelerators.  PAMELA first established that the positron
fraction rises above several GeV, rather than falling as expected for a purely
secondary component \cite{PAMELAPositron}.  AMS-02 subsequently measured the
positron fraction, the separate electron and positron fluxes, and the positron
spectral softening with much higher precision \cite{AMSPositronFraction,
AMSPositron,AMSElectron}.  The all-electron spectrum has also been measured from
space and ground by Fermi-LAT, DAMPE, CALET, and H.E.S.S., revealing a broad
softening around the TeV scale and possible additional high-energy structure
\cite{FermiCRE,DAMPEElectron,CALETElectron,HESSElectron}.  These observations
show that the high-energy lepton flux cannot be interpreted as a featureless
secondary background.

Two broad classes of explanations have been discussed.  Particle dark matter can
produce primary positrons, but is constrained by gamma rays, antiprotons, and
the smoothness of the observed spectra \cite{BergstromPositron,CirelliPPPC}.
Astrophysical explanations instead invoke primary
electron--positron pairs from pulsars and pulsar-wind nebulae, or more general
nearby discrete sources \cite{Atoyan1995,HooperPulsar,Yuksel2009,
Profumo2012,Malyshev2009,Grasso2009,DiMauro2014,Manconi2017,Fang2018}.
Mature pulsars are especially attractive: their rotational energy reservoir is
large, pair cascades are expected, and nearby systems such as Geminga and
Monogem have the right ages and distances to affect the TeV-scale lepton flux.
This has motivated many one-source or few-source interpretations of the
positron excess and of the all-electron cutoff.

There are, however, two important complications.  First, high-energy electrons
and positrons cool rapidly by synchrotron radiation and inverse-Compton
scattering, so their propagation horizon depends strongly on energy.  TeV
particles can only survive from young or nearby sources, whereas
$10$--$100\,{\rm GeV}$ particles can sample a much larger Galactic volume.
Second, arrival-direction anisotropy limits from Fermi-LAT and AMS-02 constrain
models in which a single nearby source dominates too broadly in energy
\cite{FermiAnisotropy,AMSAnisotropy}.  A collective contribution from many
unresolved pulsars would naturally reduce source-to-source fluctuations and
anisotropy compared with a single dominant object.

The idea that the lepton flux can contain both local discrete sources and a
more distant population component is not new.  Delahaye et al. treated known
local objects together with a smooth distribution beyond $2\,{\rm kpc}$ and
found that low-energy electrons can be reproduced by distant sources
\cite{Delahaye2010}.  Later population studies of pulsars and pulsar-wind
nebulae reached similar conclusions for positrons, including simulations of
catalogued and unresolved sources and fits to the AMS-02 positron fraction
\cite{Cholis2018,Manconi2020,BitterHooper2022,Orusa2021}.  These works also
emphasize that the flux above a few hundred GeV becomes increasingly sensitive
to a small number of nearby objects.  Our aim is therefore not to introduce the
collective-source idea itself, but to provide a stable two-zone electron kernel
for the middle-distance regime and to quantify how the geometric source-count
factor behaves when near-source slow diffusion is included.

The transport environment around pulsars is also not spatially uniform.
HAWC observations of extended TeV emission around Geminga and Monogem imply
slow diffusion in pulsar halos if interpreted in an isotropic diffusion picture
\cite{HAWCGeminga}.  Related slow-transport interpretations and alternatives,
including implications for the Geminga contribution to the positron excess,
have been discussed for HAWC, LHAASO, and other extended gamma-ray sources
\cite{Linden2017,HooperLinden2018,Profumo2018,Recchia2021,Zhou2022GemingaPositron,LHAASOPulsarHalo}.
Theoretical work on self-generated turbulence and near-source confinement also
suggests that particles may spend a significant time in a slow region before
joining the average interstellar medium \cite{Evoli2018,Nava2016,Fang2019}.
These results motivate a two-zone description: particles first propagate through
a slow bubble surrounding the pulsar and then through the faster interstellar
medium.

The question addressed here is whether the local positron and electron flux must
be dominated by the nearest pulsars, or whether more distant pulsars can
contribute collectively.  For one source the flux decreases with distance, but
for a population the number of sources grows with volume or disk area.  In a
three-dimensional uniform distribution the shell factor is $4\pi r^2dr$; in a
thin disk it is $2\pi R\,dR$.  The relative importance of nearby and distant
sources therefore depends on the product of the propagation kernel and the
source-count factor.  This is the quantity we calculate.

The innermost few hundred parsecs require separate care.  Pulsars trace
core-collapse supernovae and therefore massive-star formation, with progenitor
masses typically above about $8\,M_\odot$ \cite{Smartt2009CCSN}.  The Sun is
inside the Local Bubble, a hot low-density cavity produced by recent nearby
supernova activity and bounded by denser star-forming material
\cite{Lallement2014LocalBubble,Zucker2022LocalBubble}.  This environment makes
a smooth continuous pulsar surface density inside $\sim200$--$300\,{\rm pc}$
physically suspect: the region is small, currently poor in dense molecular gas,
and therefore should not be assigned the same recent birth rate as the larger
Galactic disk.  A second, independent reason is that the present pulsar
position is not generally its birth position.  Natal kick velocities are
commonly $200$--$500\,{\rm km\,s^{-1}}$ \cite{Hobbs2005Kicks}; even
$100\,{\rm km\,s^{-1}}$ moves a neutron star by about $100\,{\rm pc}$ in
$1\,{\rm Myr}$, while $10\,{\rm Myr}$ corresponds to kiloparsec-scale travel.
A neutron star born during the $10$--$20\,{\rm Myr}$ formation history of the
Local Bubble would therefore usually have left the bubble long ago, while a
neutron star observed inside it today need not have formed there.  We therefore
use the continuous disk only as a geometry diagnostic and, when comparing with
positron data, impose an empty $R<0.1\,{\rm kpc}$ cavity before allowing a
local $0.1$--$0.3\,{\rm kpc}$ annulus.

We formulate the problem with a semi-analytical two-zone electron propagation
kernel.  The source is surrounded by a slow-diffusion region of radius $R_1$,
outside which particles propagate with the interstellar diffusion coefficient.
The model is idealized, but it isolates the effect that matters here: delayed
escape from the pulsar halo combined with radiative cooling during subsequent
interstellar propagation.  The series solution is fast and stable in the
middle-distance regime where direct oscillatory integral evaluations become
poorly conditioned and where brute-force numerical transport must resolve both a
small slow bubble and a kiloparsec-scale propagation volume.

The logic of the paper is as follows.  Section~\ref{sec:model} summarizes the
electron two-zone solution.  Section~\ref{sec:data_comparison} compares
spin-down injected pulsar templates with the positron spectrum, the electron
spectrum, and all-electron anisotropy measurements.  Section~\ref{sec:discussion}
discusses the physical interpretation and the information still needed for a
complete population inference.  The finite-volume and integral-kernel checks,
the population-geometry diagnostic, and the cutoff-anisotropy diagnostic are
collected in the Appendices.

\section{Two-zone electron kernel}
\label{sec:model}

We consider electrons and positrons injected at the center of a spherical
two-zone diffusion region.  The density $N_i(E,r,t)$ in zone $i=1,2$ obeys
\begin{align}
  \frac{\partial N_i}{\partial t}
  ={}&
  D_i(E)\frac{1}{r^2}\frac{\partial}{\partial r}
  \left(r^2\frac{\partial N_i}{\partial r}\right)
  +\frac{\partial}{\partial E}\left[b(E)N_i\right]
  \nonumber\\
  &+Q(E,t)\frac{\delta(r)}{4\pi r^2}\delta_{i1}.
  \label{eq:transport}
\end{align}
The diffusion coefficient is $D_1(E)$ for $r<R_1$ and $D_2(E)$ for
$R_1<r<R_2$.  At $R_1$ the density and diffusive flux are continuous,
\begin{align}
  N_1(R_1,E,t)&=N_2(R_1,E,t),
  \nonumber\\
  D_1(E)\partial_r N_1(R_1,E,t)&=D_2(E)\partial_r N_2(R_1,E,t),
  \label{eq:interface}
\end{align}
and the outer boundary is absorbing, $N_2(R_2,E,t)=0$.  In the applications
below we use $R_2=20\,{\rm kpc}=2\times10^4\,{\rm pc}$, chosen large enough
that it acts only as a numerical boundary.

The key simplification is that $D_1$ and $D_2$ have the same energy dependence,
\begin{align}
  D_1(E)&=D_{10} f_D(E),
  &
  D_2(E)&=D_{20} f_D(E),
  \nonumber\\
  \eta&\equiv\frac{D_2}{D_1}={\rm const}.
  \label{eq:constant_eta}
\end{align}
The spatial eigenfunctions are then the same as in the fixed-energy proton
problem \cite{2026arXiv260619701L}.  Defining
\begin{equation}
  \beta=\frac{R_2}{R_1},\qquad
  m=\frac{\beta-1}{\sqrt{\eta}},
  \label{eq:beta_m}
\end{equation}
the eigenvalues $x_n>0$ are roots of
\begin{equation}
  \sqrt{\eta}\,x\sin x\cos(mx)
  +\left[x\cos x+(\eta-1)\sin x\right]\sin(mx)=0 .
  \label{eq:roots}
\end{equation}
The roots are found with the Chebyshev-polynomial bracketing method used in
the accompanying code.

Radiative cooling enters through the characteristics.  For a burst source of
age $T$, the injection energy $E_s$ is determined by
\begin{equation}
  T=\int_E^{E_s}\frac{dE'}{b(E')}.
  \label{eq:source_energy}
\end{equation}
No contribution exists if this equation has no finite solution.  The modal
diffusion variable is
\begin{align}
  \Lambda_1(E,E_s)=
  \int_E^{E_s}\frac{D_1(E')}{b(E')}\,dE',
  \nonumber\\
  S_n(E,E_s)=\left(\frac{x_n}{R_1}\right)^2\Lambda_1(E,E_s).
  \label{eq:lambda}
\end{align}

For each root we define
\begin{equation}
  \theta_n=\frac{x_n(\beta-1)}{\sqrt{\eta}} .
  \label{eq:theta_n}
\end{equation}
The radial eigenfunctions are written as
\begin{align}
  \psi_{1n}(r)&=\frac{\sin(x_nr/R_1)}{r},
  \qquad 0<r<R_1,
  \nonumber\\
  \psi_{2n}(r)&=
  B_n\frac{\sin[x_n(\beta-r/R_1)/\sqrt{\eta}]}{r},
  \qquad R_1<r<R_2,
  \label{eq:eigenfunctions}
\end{align}
with the interface amplitude
\begin{equation}
  B_n=\frac{\sin x_n}{\sin\theta_n}.
  \label{eq:Bn}
\end{equation}
The eigenvalue equation~\eqref{eq:roots} is precisely the remaining
condition from continuity of $D\partial_r\psi_n$ at $R_1$ after imposing
Eq.~\eqref{eq:Bn}.  The normalization used below is
\begin{align}
  I_n
  &=
  \int_0^{R_1}dr\,r^2\psi_{1n}^2
  +\int_{R_1}^{R_2}dr\,r^2\psi_{2n}^2
  \nonumber\\
  &=
  \frac{R_1}{2}
  \left(1-\frac{\sin2x_n}{2x_n}\right)
  \nonumber\\
  &\quad
  +R_1B_n^2
  \left[
  \frac{\beta-1}{2}
  -\frac{\sqrt{\eta}\sin2\theta_n}{4x_n}
  \right].
  \label{eq:In}
\end{align}

The burst solution in the outer region, relevant for the Earth when the source
is outside its own slow zone, is then
\begin{align}
  N_2(E,r,T)
  ={}&
  \frac{b(E_s)}{b(E)}
  \frac{Q(E_s)}{4\pi R_1}
  \sum_{n=1}^{\infty}
  \frac{x_n}{I_n}
  B_n
  \nonumber\\
  &\times
  \frac{\sin\!\left[
    x_n(\beta-r/R_1)/\sqrt{\eta}
  \right]}{r}
  e^{-S_n(E,E_s)} ,
  \label{eq:outer_solution}
\end{align}
The inner-zone expression is obtained from the same formula by replacing the
last spatial factor with $\sin(x_nr/R_1)/r$.  This form is algebraically
equivalent to the derivative-of-characteristic normalization; it is numerically
convenient because the normalization in Eq.~\eqref{eq:In} is positive definite.

The loss function includes synchrotron cooling and inverse-Compton losses on
blackbody radiation fields with Klein--Nishina suppression,
\begin{equation}
  b(E)=b_{\rm syn}(E)+\sum_j b_{{\rm IC},j}(E),
  \label{eq:loss}
\end{equation}
with CMB, infrared, and optical components.  Each radiation component is
treated as a diluted blackbody with temperature $T_j$ and energy density $U_j$,
\begin{align}
  b_{{\rm IC},j}(E)
  ={}&
  \frac{4}{3}\sigma_T c\,U_j\,\gamma^2
  F_{\rm KN}(u_j),
  \nonumber\\
  u_j={}&4\gamma\Theta_j,\qquad
  \Theta_j=\frac{k_BT_j}{m_ec^2},
  \label{eq:ic_blackbody_loss}
\end{align}
where $F_{\rm KN}$ is the analytic blackbody Klein--Nishina correction.  We use
the Khangulyan--Aharonian--Kelner approximation \cite{Khangulyan}, normalized so that
$F_{\rm KN}\rightarrow1$ in the Thomson limit:
\begin{equation}
  F_{\rm KN}(u)=
  \frac{F_{\rm iso}(u)}{(\pi^4/135)u^2},
  \label{eq:fkn}
\end{equation}
with
\begin{align}
  F_{\rm iso}(u)
  ={}&
  \frac{c_{\rm iso}u
  \ln\!\left(1+0.722u/c_{\rm iso}\right)}
  {1+c_{\rm iso}u/0.822}
  \nonumber\\
  &\times
  \left[
  1+\frac{a_{\rm iso}u^{\alpha_{\rm iso}}}
  {1+b_{\rm iso}u^{\beta_{\rm iso}}}
  \right]^{-1}.
  \label{eq:fiso}
\end{align}
The constants are
$c_{\rm iso}=5.68$, $a_{\rm iso}=-0.362$, $\alpha_{\rm iso}=0.682$,
$b_{\rm iso}=0.826$, and $\beta_{\rm iso}=1.281$.  The same blackbody
components should be used for the electron cooling and for any subsequent
inverse-Compton photon calculation.

\section{Flux and anisotropy comparison}
\label{sec:data_comparison}

We now connect the propagation diagnostic to the measured positron flux.  This
step is not a full Galactic population fit; it is a controlled data comparison
designed to test whether the conclusion survives three effects that are absent
from the idealized shell calculation: spin-down injection, discreteness of the
nearest sources, and dipole anisotropy constraints.

\subsection{Spin-down injection and the local boundary}
\label{sec:spindown_local}

For the data comparison we use a spin-down luminosity history
\begin{equation}
  L(t)=L_0\left(1+\frac{t}{\tau_0}\right)^{-2},
  \qquad \tau_0=10\,{\rm kyr},
  \label{eq:spindown_luminosity}
\end{equation}
and remove the first $t_{\rm delay}=10\,{\rm kyr}$ of injected pairs.  This
delay is a simple phenomenological way to exclude the early supernova-remnant
reverberation phase, during which pairs are not expected to escape into the
interstellar medium as freely as in the mature pulsar-halo stage.  For a
catalogued pulsar of characteristic age $T$ and present spin-down power
$\dot E$, the luminosity at emission time $t_{\rm em}$ is written as
\begin{equation}
  L(t_{\rm em})=\dot E
  \left(\frac{T+\tau_0}{t_{\rm em}+\tau_0}\right)^2,
  \qquad
  t_{\rm delay}<t_{\rm em}<T .
  \label{eq:spindown_catalog_luminosity}
\end{equation}

The Solar neighborhood is not well represented by an unbroken continuous pulsar
disk.  We therefore remove newly born sources inside $R<0.1\,{\rm kpc}$ in
the population fit.  This encodes the Local-Bubble and proper-motion arguments
that no very nearby young pulsar population should be inserted by hand.  To
avoid forcing the pulsar templates to fit the entire positron spectrum, the
fit profiles two additional non-negative components: a steep secondary-like
positron basis and a smooth power-law background.  The pulsar templates
themselves use a broken-power-law pair injection spectrum, as commonly adopted
in time-dependent PWN modeling \cite{Torres2014PWN}; here we use the best-fit
population-MCMC shape, with $E_b=509\,{\rm GeV}$.

\begin{figure}[t]
\includegraphics[width=\linewidth]{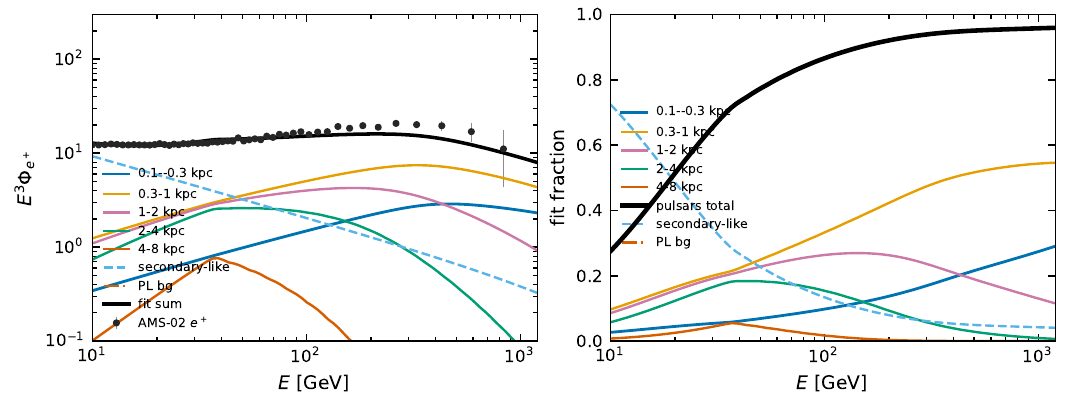}
\caption{Fit to the AMS-02 positron flux used in the population comparison.
The black curve is the profiled total model.  The blue curve is the continuous
spin-down pulsar-pair component after removing newly born sources inside
$R<0.1\,{\rm kpc}$; the orange curve is the secondary-like basis.  The dotted
curves split the same continuous-disk pulsar kernel into the local
$0.1$--$0.3\,{\rm kpc}$ annulus and the $R>0.3\,{\rm kpc}$ component.  The
$R>0.3\,{\rm kpc}$ component dominates the fitted pulsar contribution over the
plotted energy range.}
\label{fig:positron_fit}
\end{figure}

Table~\ref{tab:near_far_annuli} and Fig.~\ref{fig:positron_fit} decompose
the fitted continuous pulsar population by distance annulus.  They use the
same continuous spin-down injection, $10\,{\rm kyr}$ release delay, common
two-zone halo, secondary-like component, and power-law background as the MCMC
fit to the AMS-02 2021 positron flux \cite{AMS2021PhysicsReports}.  The
annuli share one pulsar-density normalization, so the plotted curves are a
decomposition of a single fitted population rather than independently rescaled
templates.  The continuous spatial distribution is only a source-count
approximation: each source is first propagated with its own local spherical
two-zone kernel, and the annulus integral then replaces the discrete source
sum.  We therefore neglect halo overlap, collective changes to the diffusion
environment, and the detailed Galactic boundary geometry.  The total flux
below tens of GeV is mostly assigned to the
secondary-like basis, but within the fitted pulsar component the
$R>0.3\,{\rm kpc}$ annuli dominate over the local $0.1$--$0.3\,{\rm kpc}$
annulus at all plotted energies.  This dominance is a source-count effect, not
an enhancement of distant propagation by the slow halo.  The one-zone reference
problem assumes that pairs propagate with $D_{\rm ISM}(E)$ immediately after
injection.  The two-zone problem instead inserts a slow near-source residence
time before the particles enter the interstellar zone.  For electrons and
positrons this additional residence time reduces the absolute flux, because the
particles cool while still inside the slow halo.  For the same disk geometry
and injection history, replacing the two-zone kernel by a one-zone interstellar
kernel gives a larger absolute distant flux: the two-zone
$R>0.3\,{\rm kpc}$ contribution is smaller by factors
$0.97$, $0.97$, $0.94$, $0.85$, and $0.67$ at
$10$, $30$, $100$, $300$, and $800\,{\rm GeV}$, respectively.  The distant
fraction changes much less, because both local and distant annuli share the
same near-source escape delay.  Thus the main effect of the two-zone kernel is
not to make distant sources brighter, but to introduce an energy-dependent
cooling penalty that becomes visible toward the high-energy cutoff.

\begin{table*}[t]
\caption{Best-fit fractions in the continuous spin-down disk-population
decomposition after removing newly born sources inside $R<0.1\,{\rm kpc}$.
The pulsar columns are distance-bin contributions of the fitted continuous
population.  The secondary-like and power-law background components are
profiled simultaneously.  All entries are fractions of the total fitted
positron flux.}
\label{tab:near_far_annuli}
\begin{ruledtabular}
\begin{tabular}{cccccc}
$E$ [GeV] & $0.1$--$0.3$ & $R>0.3$ pulsars &
pulsars total & secondary-like & PL bg \\
\hline
10  & 0.027 & 0.249 & 0.276 & 0.724 & 0.000 \\
29  & 0.054 & 0.592 & 0.646 & 0.354 & 0.000 \\
101 & 0.098 & 0.768 & 0.866 & 0.134 & 0.000 \\
294 & 0.168 & 0.770 & 0.938 & 0.062 & 0.000 \\
487 & 0.213 & 0.737 & 0.950 & 0.050 & 0.000 \\
809 & 0.256 & 0.700 & 0.956 & 0.044 & 0.000
\end{tabular}
\end{ruledtabular}
\end{table*}

\subsection{Continuous spin-down common-halo MCMC}
\label{sec:common_halo}

We next perform a solver-in-the-loop MCMC in which all catalogued pulsars with
$0.1\leq d\leq2\,{\rm kpc}$, characteristic age $T\leq3\,{\rm Myr}$, and
finite $\dot E$ share the same slow-diffusion halo parameters $R_1$ and
$D_1$.  The innermost $0.1\,{\rm kpc}$ is removed for the same Local-Bubble
and proper-motion reasons discussed above.  Each source is treated as a
continuous spin-down injector after a fixed
$t_{\rm delay}=10\,{\rm kyr}$ release time, rather than as an impulsive burst.
The pair injection spectrum is
\begin{equation}
  Q(E,t)\propto L(t)
  \begin{cases}
    (E/E_b)^{-\gamma_1}, & E<E_b,\\
    (E/E_b)^{-\gamma_2}, & E\geq E_b,
  \end{cases}
  \exp(-E/E_{\rm cut}),
  \label{eq:pwn_bpl_injection}
\end{equation}
with $E_{\rm cut}=100\,{\rm TeV}$.  The normalization is tied to the
instantaneous spin-down luminosity in Eq.~\eqref{eq:spindown_luminosity};
the reported efficiency scale is the profiled non-negative multiplier of the
catalogued pulsar template.  The
interstellar diffusion coefficient is fixed to
\begin{equation}
  D_2(E)=3\times10^{28}
  \left(\frac{E}{\rm GeV}\right)^{1/3}
  {\rm cm^2\,s^{-1}},
  \label{eq:dism_fixed}
\end{equation}
with the energy index fixed to $1/3$.

The sampled nonlinear parameters are
$\{R_1,\log_{10}D_1(100\,{\rm TeV}),\gamma_1,\gamma_2,
\log_{10}E_b,\alpha_{\rm bg}\}$.  At every MCMC point we profile over three
non-negative linear amplitudes: the catalogued-pulsar positron template, a
steep secondary-like positron basis
$\Phi_{\rm sec}\propto E^{-3.65}\exp(-E/5\,{\rm TeV})$, and a smooth
power-law background
$\Phi_{\rm bg}\propto E^{-\alpha_{\rm bg}}$.  This is still not a complete
Galactic secondary calculation, but it avoids forcing the pulsar template to
absorb the conventional positron background.  In particular, the
secondary-like basis is not computed from a matched proton spectrum, diffusion
index, gas distribution, and inclusive production cross section; it is only a
flexible spectral component for profiling the low-energy positron background.
A fully matched secondary calculation is possible, but its normalization and
spectral curvature would be strongly degenerate with the pulsar efficiency,
injection slopes, release history, and halo parameters in the present fit.
We therefore treat the secondary contribution as a nuisance background rather
than as an independently inferred Galactic propagation component.

For anisotropy we use the vector flux-weighted dipole
\begin{equation}
  \boldsymbol{\Delta}(E)
  =
  \frac{1}{\Phi_{\rm CRE}(E)}
  \sum_i
  \Delta_i(E)\Phi_i(E)\,\hat{\boldsymbol n}_i ,
  \label{eq:anis_vector_sum}
\end{equation}
where $\Phi_{\rm CRE}$ is the measured all-electron plus positron flux used as
the denominator and $\hat{\boldsymbol n}_i$ points from the source to the
observer.  For a single source this is evaluated from the two-zone density
gradient at the observer,
\begin{equation}
  \Delta_i(E)
  \simeq
  \frac{3D_{\rm obs}(E)}{c}
  \left|
  \frac{\partial_r N_i(E,r)}{N_i(E,r)}
  \right|_{r=d_i},
  \label{eq:single_source_dipole}
\end{equation}
with $D_{\rm obs}=D_2$ when the observer lies outside the source halo.  This
is the standard diffusion dipole expression applied to the two-zone Green
function; the simplification assumes that the local gradient is dominated by
the radial source-observer direction and that the measured all-electron flux
provides the isotropic denominator.  The fitted anisotropy data are the CALET
Analysis-B fixed-bin
dipole amplitudes \cite{CALETAnisotropy2025}, which are independent energy
bins following the Fermi-LAT binning with an additional $1$--$5\,{\rm TeV}$
bin.  Since the published CALET figure gives amplitudes and confidence limits
rather than a machine-readable vector likelihood, we use an amplitude-only
approximation.  If the three Cartesian components of the reconstructed dipole
have independent Gaussian noise with common width $\sigma_j$ in bin $j$, then
the measured amplitude $\delta_j$ has a noncentral-chi distribution with three
degrees of freedom,
\begin{equation}
  p(\delta_j|\Delta_j,\sigma_j)
  =
  \frac{2\delta_j}{\sigma_j^2}
  f_{\chi^2_3(\lambda_j)}
  \left(\frac{\delta_j^2}{\sigma_j^2}\right),
  \qquad
  \lambda_j=\frac{\Delta_j^2}{\sigma_j^2},
  \label{eq:calet_anis_likelihood}
\end{equation}
where $\Delta_j$ is the model amplitude and $f_{\chi^2_3(\lambda)}$ is the
noncentral $\chi^2$ density.  We calibrate $\sigma_j$ separately in each bin
by requiring the profile-likelihood point at the digitized CALET frequentist
95\% upper limit to satisfy
$2[\ln p(\delta_j|\hat\Delta_j,\sigma_j)-
\ln p(\delta_j|\Delta_{95,j},\sigma_j)]=3.84$.
The anisotropy contribution to the sampled objective is therefore
\begin{equation}
  \chi^2_{\rm CALET}
  =
  -2\sum_j
  \ln\frac{p(\delta_j|\Delta_j,\sigma_j)}
          {p(\delta_j|\hat\Delta_j,\sigma_j)} .
  \label{eq:calet_anis_chi2}
\end{equation}
This uses the measured CALET amplitudes rather than only their upper limits,
but it is still an approximate amplitude-only likelihood.  We also keep a
smooth approximation to the Fermi-LAT 2017 all-electron dipole upper limits,
anchored at $3\times10^{-3}$ near $42\,{\rm GeV}$ and $3\times10^{-2}$ near
$2\,{\rm TeV}$, as a conservative one-sided consistency penalty.

The sampled objective is
$\exp[-(\chi^2_{e^+}+\chi^2_{\rm CALET}+\chi^2_{\rm UL})/2]$, where
$\chi^2_{e^+}$ is the AMS-02 positron Gaussian chi-square after profiling
over the three linear amplitudes, $\chi^2_{\rm CALET}$ is given by
Eq.~\eqref{eq:calet_anis_chi2}, and $\chi^2_{\rm UL}$ is the one-sided
Fermi-LAT consistency penalty.  The production chain used 64 walkers for
900 steps on 80 CPU cores, with 250 burn-in steps, $n_{\rm root}=700$, a
4096-point cooling table, and 16 logarithmic time-integration points per
source.  We retain 5000 posterior samples for the figures and intervals.

The best sampled point has
\begin{equation}
\begin{gathered}
  R_1=0.0683\,{\rm kpc},\\
  D_1(100\,{\rm TeV})=8.85\times10^{27}\,{\rm cm^2\,s^{-1}},\\
  \gamma_1=2.00,\quad
  \gamma_2=2.61,\quad
  E_b=509\,{\rm GeV},\\
  \alpha_{\rm bg}=3.21,\quad
  \chi^2_{e^+}/{\rm dof}=0.590 .
\end{gathered}
\label{eq:best_population_mcmc}
\end{equation}
The same point has $\chi^2_{\rm CALET}=9.7\times10^{-3}$, no Fermi-LAT
upper-limit penalty, a maximum Fermi diagnostic ratio
$\Delta/\Delta_{\rm UL}=0.55$, and a profiled pulsar efficiency scale
$0.160$.  The corresponding diffusion contrast is $\eta=157$ at the reference
energy $1\,{\rm GeV}$.  The posterior medians and 16--84 percentile intervals
are
\begin{align}
  R_1&=0.0709^{+0.0510}_{-0.0407}\,{\rm kpc},
  \nonumber\\
  D_1(100\,{\rm TeV})
  &=2.08^{+4.97}_{-1.65}\times10^{28}\,{\rm cm^2\,s^{-1}},
  \nonumber\\
  \gamma_1&=1.86^{+0.17}_{-0.37},\qquad
  \gamma_2=2.66^{+0.35}_{-0.35},
  \nonumber\\
  E_b&=326^{+253}_{-157}\,{\rm GeV},\qquad
  \alpha_{\rm bg}=2.97^{+0.19}_{-0.13}.
  \label{eq:population_mcmc_intervals}
\end{align}
The median values of the diagnostic quantities are
$\chi^2_{e^+}/{\rm dof}=0.661$, $\chi^2_{\rm CALET}=3.4\times10^{-3}$, and
$\max(\Delta/\Delta_{\rm UL})=0.35$.

\begin{table*}[t]
\caption{Continuous spin-down common-halo MCMC results.  The interval column
lists the 16, 50, and 84 percentiles of the retained posterior samples.  The
linear amplitudes of the pulsar, secondary-like, and smooth-background
templates are profiled at every sampled point and are therefore not listed as
sampled parameters.}
\label{tab:population_mcmc}
\begin{ruledtabular}
\begin{tabular}{cccc}
Parameter & Prior range & 16--50--84 percentile & Best point \\
\hline
$R_1$ [pc] & $10$--$150$ & $30.3$--$70.9$--$121.9$ & $68.3$ \\
$\log_{10}D_1(100\,{\rm TeV})$ & $26.5$--$29.1$ &
$27.64$--$28.32$--$28.85$ & $27.95$ \\
$\gamma_1$ & $0.8$--$2.1$ & $1.49$--$1.86$--$2.04$ & $2.00$ \\
$\gamma_2$ & $1.6$--$3.2$ & $2.31$--$2.66$--$3.01$ & $2.61$ \\
$E_b$ [GeV] & $30$--$2000$ & $169$--$326$--$579$ & $509$ \\
$\alpha_{\rm bg}$ & $2.7$--$4.5$ & $2.84$--$2.97$--$3.16$ & $3.21$
\end{tabular}
\end{ruledtabular}
\end{table*}

\begin{figure}[t]
\includegraphics[width=\linewidth]{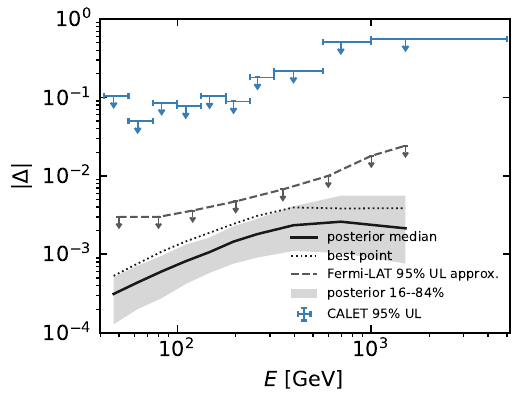}
\caption{All-electron dipole anisotropy constraints.  Downward arrows mark
95\% upper limits: the blue points show the CALET fixed-bin limits, with
horizontal bars indicating the energy bins, and the gray dashed curve with
arrows shows the smooth Fermi-LAT upper-limit approximation used as a
one-sided consistency penalty.  The solid black curve and gray band show the
posterior median and 16--84 percentile prediction for the model dipole
amplitude, while the dotted curve marks the best sampled point.}
\label{fig:anisotropy_fit}
\end{figure}

Figure~\ref{fig:anisotropy_fit} shows the anisotropy constraints in the
standard cosmic-ray format: dipole amplitude as a function of energy, with
horizontal bars marking the energy bins.  We plot the CALET fixed-bin 95\%
upper limits, the smooth Fermi-LAT upper-limit approximation, and the
posterior prediction for the model dipole amplitude.  The shaded band is the
16--84 percentile interval obtained by propagating retained population-MCMC
samples through the anisotropy calculation; it is a model-prediction band, not
a refit of the CALET or Fermi-LAT upper-limit points.  The CALET measured
fixed-bin amplitudes are not shown as constraints in this figure, because they
are dominated by finite-statistics white-noise amplitudes and are therefore
not directly comparable to the Fermi-LAT upper-limit curve.

This result revises the interpretation of the earlier two-parameter diagnostic
scan.  With fixed injection indices and a single smooth background, the flux
fit preferred an effective compact radius near $R_1\simeq20\,{\rm pc}$.  Once
continuous spin-down injection, a broken-power-law injection shape, and
separate secondary-like and smooth background templates are included, the
posterior moves to $R_1\simeq70\,{\rm pc}$ with a broad allowed range.  This is
comparable to the slow-diffusion scales often discussed for Geminga and
Monogem in TeV gamma-ray morphology studies
\cite{HAWCGeminga,Recchia2021,Schroer2023}.  The shift shows that the
compact-radius result is a conditional effective preference of the restrictive
fit, not a robust measurement of a physical subdiffusion halo.  Conversely, the
new MCMC result should also not be overinterpreted as a direct halo-size
measurement: positron flux and amplitude-only anisotropy data still leave
degeneracies with injection shape, background modeling, release history, and
catalog incompleteness.

\begin{figure}[t]
\includegraphics[width=\linewidth]{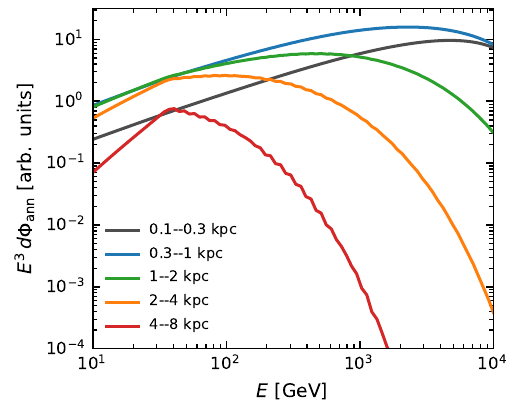}
\caption{Age-integrated electron/positron contribution from a continuous
finite-thickness disk distribution of impulsive sources.  The source age, or
diffusion time after the impulsive release, is integrated continuously from
$0.03$ to $10\,{\rm Myr}$ with a uniform birth-rate weight.  The empty
$R<0.1\,{\rm kpc}$ Local-Bubble cavity is removed; the first populated local
annulus is $0.1$--$0.3\,{\rm kpc}$.  All annuli have the same source density
and $h=0.2\,{\rm kpc}$.}
\label{fig:annulus_flux}
\end{figure}

The annulus decomposition in Fig.~\ref{fig:annulus_flux} gives the same physical
message in flux space.  It uses an impulsive continuous-disk diagnostic rather
than the continuous spin-down fit, but it makes the geometric point transparent:
once the inner $R<0.1\,{\rm kpc}$ cavity is removed, annuli beyond
$0.3\,{\rm kpc}$ remain important at tens to hundreds of GeV, while the TeV
contribution shifts inward.  A one-zone propagation kernel with the same
interstellar diffusion coefficient gives slightly larger absolute annulus
fluxes, especially near the cooling cutoff, but it does not qualitatively
change this geometric ordering.

\section{Discussion}
\label{sec:discussion}

The geometry diagnostic implies a two-component interpretation of the local
lepton flux.  In a three-dimensional uniform population, the volume factor is
strong enough that $10$--$100\,{\rm GeV}$ electrons and positrons are naturally
dominated by kiloparsec-scale shells.  In a disk population the effect is
weaker, because the source count grows as $R\,dR$ rather than $r^2dr$.
Nevertheless, the disk calculation still leaves a substantial fraction of the
sub-TeV flux outside $1\,{\rm kpc}$.  In this regime a one-source
interpretation is not forced by propagation.  A large number of unresolved
sources can reduce stochastic fluctuations and can produce a smoother spectrum
and a smaller dipole anisotropy than a dominant nearby source.

The comparison with a one-zone propagation kernel clarifies the role of the
slow halo.  In a one-zone model the elapsed time is controlled only by
interstellar diffusion from the source to the observer.  In the two-zone model
the particle first spends an additional escape time in the inhibited-diffusion
region.  This extra time affects all annuli similarly at low energy, where
radiative losses are weak, so the geometric ordering of near and distant
annuli is almost unchanged.  At higher energy the same delay becomes a cooling
filter and suppresses distant annuli more strongly.  The two-zone result should
therefore be read as a conservative propagation kernel for distant pulsar
contributions, rather than as a mechanism that boosts them relative to a
one-zone ISM calculation.

The comparison with AMS-02 positrons sharpens this statement.  Once the
innermost $100\,{\rm pc}$ is removed, the fitted continuous pulsar population
is dominated by the $R>0.3\,{\rm kpc}$ annuli, not by the local
$0.1$--$0.3\,{\rm kpc}$ annulus.  In Table~\ref{tab:near_far_annuli}, the
$R>0.3\,{\rm kpc}$ pulsar contribution is about an order of magnitude larger
than the local annulus near $10\,{\rm GeV}$ and remains larger through the
sub-TeV range.  The total positron flux at the lowest energies is still mostly
assigned to the secondary-like basis, so ``$300\,{\rm pc}+$ dominated'' should
be understood as a statement about the fitted pulsar component, not about the
entire positron flux.  This is consistent with the Local-Bubble picture: a
smooth recent birth rate of pulsars inside the nearest few hundred parsecs is
not a good physical prior, while the larger disk area beyond $0.3\,{\rm kpc}$
naturally supplies most of the continuous pulsar contribution.

Near the TeV scale radiative cooling restricts the source age and distance that
can contribute.  The cumulative shell contribution then shifts inward, making
nearby mature pulsars such as Geminga and Monogem much more relevant.  This is
also the regime where a single source can imprint a cutoff-like spectral
feature and potentially a measurable dipole anisotropy.  At these energies the
continuous-source approximation should be viewed as an ensemble mean: the
actual flux and dipole can fluctuate stochastically with the local realization
of source ages, distances, luminosities, and proper motions.  Similar
multiple-population and local-source effects have also been emphasized in
recent fits to Galactic cosmic-ray nuclei \cite{Yuan2026MultipleSources}.  The
common-halo MCMC shows that anisotropy is not a passive afterthought: the allowed region
must fit the AMS-02 positron shape while remaining compatible with the CALET
measured amplitudes and the approximate Fermi-LAT upper limits.  With
continuous spin-down injection and flexible spectral/background shapes, this
condition is satisfied for a broad posterior centered near
$R_1\simeq70\,{\rm pc}$ rather than for the compact
$R_1\simeq20\,{\rm pc}$ radius found in the more restrictive diagnostic scan.

The calculation here remains deliberately limited.  It includes a
solver-in-the-loop MCMC over the common-halo parameters and several spectral
shape parameters, but not a full posterior over a realistic Galactic pulsar
population.  It assumes a common halo for all catalogued sources in the
anisotropy comparison and uses an amplitude-only approximation to the CALET
dipole likelihood rather than the official full vector likelihood.  The
secondary-like positron term is a flexible basis rather than a propagation
calculation of secondaries and is therefore not guaranteed to match a chosen
primary proton spectrum, diffusion index, gas map, or production cross
section.  We do not include such a calculation here because it would introduce
additional propagation and cross-section nuisance parameters that are strongly
degenerate with the pulsar population parameters being tested.  A complete data
analysis should include a physically computed secondary positron background, a
smooth primary electron component, a luminosity and age distribution for
pulsars, spiral-arm geometry, catalog incompleteness, distance and age
uncertainties, stochastic realizations of the nearby high-energy sources, and
the exact AMS-02, Fermi-LAT, and CALET anisotropy likelihoods.  The present
result is therefore a propagation-controlled demonstration that positron flux
and anisotropy can test common-halo parameter space, but not yet a final
population inference or a unique halo-size measurement.

Additional observables are needed to turn the consistency test into a
determination of the slow-diffusion region.  The most direct information is
the TeV inverse-Compton surface-brightness profile and spectrum of individual
halos such as Geminga and Monogem, because these data probe the angular extent,
cooling history, and energy dependence of particles before they contribute to
the local lepton flux.  Population information is also important: pulsar
proper motions, birth sites, distances, ages, and gamma-ray efficiencies can
separate source physics from propagation.  Finally, the all-electron spectrum
and anisotropy direction, not only the dipole amplitude, would help distinguish
a few nearby sources from a collective nonlocal contribution.  Without these
inputs, changes in injection efficiency, release delay, background shape, and
source incompleteness can mimic changes in $R_1$ and $D_1$.

Another practical lesson concerns numerical solvers.  A finite-volume scheme is
useful for testing and for extensions beyond spherical symmetry, but a
low-order energy-loss update can artificially populate energies above the
cooling cutoff.  For parameter scans near the cutoff, the semi-analytical
series solution is more reliable.  A production finite-volume calculation
should use higher energy resolution or a characteristic/energy-remapping update
for the cooling term.

\section{Conclusions}
\label{sec:conclusions}

We developed and applied a two-zone electron series solution with radiative
cooling to the question of whether local cosmic-ray positrons and electrons
must come from nearby pulsars.  Our main conclusions are:
\begin{enumerate}
  \item For a three-dimensional uniform pulsar population, the shell factor
  $4\pi r^2dr$ makes
  kiloparsec distances dominate the age-integrated contribution at
  $10$--$100\,{\rm GeV}$.
  \item In a disk-like population with $h=0.2\,{\rm kpc}$, the contribution is
  shifted inward, but sources beyond $1\,{\rm kpc}$ still supply
  $37$--$47\%$ of the $10$--$100\,{\rm GeV}$ flux in the fiducial model.
  \item At $300\,{\rm GeV}$ the contribution begins to localize in both
  geometries.
  \item At $1\,{\rm TeV}$ radiative cooling makes the flux local:
  approximately two thirds of the contribution comes from within
  $1\,{\rm kpc}$ in the three-dimensional model and about $90\%$ in the disk
  model.
  \item With continuous spin-down injection, a $10\,{\rm kyr}$ early-release
  delay, and no newly born source inside $R<0.1\,{\rm kpc}$, the fit to AMS-02
  positrons decomposes into a pulsar fraction of about $28\%$ at
  $10\,{\rm GeV}$ and more than $85\%$ above $100\,{\rm GeV}$, with the
  remaining low-energy flux mostly assigned to the secondary-like basis.
  \item In the continuous-disk annulus decomposition of this fitted pulsar
  component, sources beyond $0.3\,{\rm kpc}$ dominate over the local
  $0.1$--$0.3\,{\rm kpc}$ annulus at all plotted energies.  The dominance is
  strongest at low energy and remains present through the sub-TeV range.
  \item Relative to a one-zone interstellar diffusion kernel, the two-zone halo
  reduces the absolute distant-source flux rather than increasing it.  For
  $R>0.3\,{\rm kpc}$ the two-zone/one-zone ratio is about $0.94$ at
  $100\,{\rm GeV}$, $0.85$ at $300\,{\rm GeV}$, and $0.67$ at
  $800\,{\rm GeV}$.  The survival of the distant component is therefore a
  population-geometry effect, not a slow-halo enhancement.
  \item A continuous spin-down common-halo MCMC with fixed
  $D_2=3\times10^{28}(E/{\rm GeV})^{1/3}\,{\rm cm^2\,s^{-1}}$ finds
  anisotropy-safe posterior support around
  $R_1=0.0709^{+0.0510}_{-0.0407}\,{\rm kpc}$ and
  $D_1(100\,{\rm TeV})
  =2.08^{+4.97}_{-1.65}\times10^{28}\,{\rm cm^2\,s^{-1}}$.
  This Geminga-scale result replaces the compact effective radius found in the
  earlier restrictive two-parameter diagnostic, but it is still a conditional
  population-model constraint rather than a unique halo-size measurement from
  flux and anisotropy alone.
\end{enumerate}

Therefore, a collective pulsar component beyond the immediate Local-Bubble
neighborhood is a plausible contributor to the positron and electron flux below
a few hundred GeV, while the TeV-scale behavior remains sensitive to the local
catalogued population and to anisotropy constraints.

\appendix

\section{Finite-volume and integral-kernel checks}
\label{sec:numerical}

We implemented an independent spherical finite-volume solver for
Eq.~\eqref{eq:transport}.  The radial diffusion operator is solved implicitly
with a tridiagonal Thomas algorithm for each energy bin, and the energy-loss
term is updated with a conservative upwind finite-volume step.  The default
grid used for the comparison has 256 radial cells and 128 logarithmic energy
bins from $1$ to $1000\,{\rm TeV}$.

\begin{figure*}[t]
\includegraphics[width=0.96\textwidth]{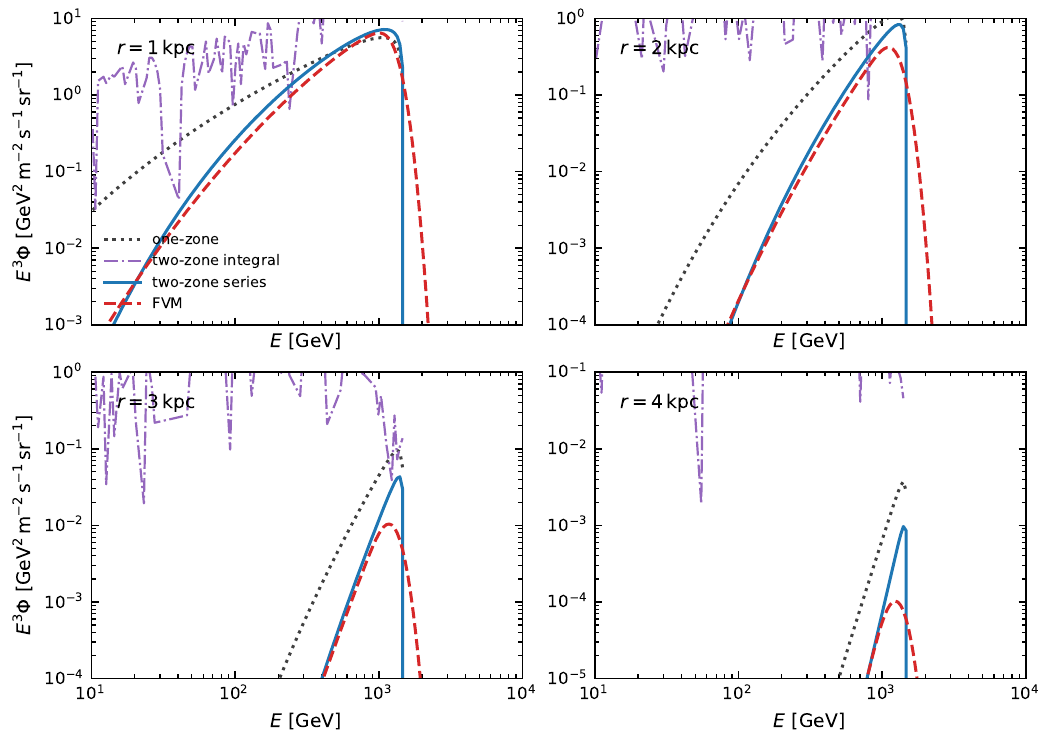}
\caption{Comparison of the finite-volume solution, the two-zone series solution,
the Osipov-type two-zone integral solution, and the one-zone solution for a
Geminga-like burst source.  The calculation uses $R_2=20\,{\rm kpc}$ for the
series and FVM calculations to remove outer-boundary effects.  The panels show
the spectra at $r=1$, $2$, $3$, and $4\,{\rm kpc}$.  The vertical range in
each panel is separately zoomed to four decades around the cutoff region where
the finite-volume and series results differ most clearly.
The Osipov-type integral curve is evaluated by direct quadrature and
illustrates the numerical ringing that appears for distant-source parameters.}
\label{fig:solution_comparison}
\end{figure*}

The finite-volume solution should not, however, be treated as a high-precision
replacement for the analytical series near a radiative cooling boundary.  To
show this explicitly, Fig.~\ref{fig:solution_comparison} compares the one-zone
solution, the Osipov-type two-zone integral solution, the two-zone series, and
the finite-volume result for a burst source of age $0.342\,{\rm Myr}$ at
intermediate and large distances.  The outer boundary is placed at
$R_2=20\,{\rm kpc}$, sufficiently large that it does not control the plotted
spectra.  The two-zone series includes the slow bubble and gives a sharp
physical cutoff when Eq.~\eqref{eq:source_energy} has no finite solution.  The
finite-volume result tracks the series over much of the spectrum but smears the
cutoff and leaves a numerical high-energy tail.  This is the main practical
advantage of the characteristic series solution for distant-source and cutoff
studies.  The direct evaluation of the Osipov-type integral is included to
show the numerical instability of the oscillatory two-zone integral kernel in
the same physical parameter range.

A useful physical regime is a Geminga-age burst source,
$T=0.342\,{\rm Myr}$, embedded in a $R_1=50\,{\rm pc}$ slow-diffusion bubble
with $D_1(100\,{\rm TeV})=4.5\times10^{27}\,{\rm cm^2\,s^{-1}}$ and
interstellar $D_2=3\times10^{28}(E/{\rm GeV})^{1/3}\,{\rm cm^2\,s^{-1}}$,
observed from $r=2$--$4\,{\rm kpc}$.  These distances are not meant to
represent Geminga itself, but are typical of the unresolved Galactic pulsar
population relevant to the collective component.  Around the cooling cutoff,
$E\simeq1.2$--$1.5\,{\rm TeV}$, the finite-volume calculation becomes
resolution limited: at $E=1.40\,{\rm TeV}$ the FVM/series density ratio is
$0.28$, $0.15$, and $0.079$ for $r=2$, $3$, and $4\,{\rm kpc}$, respectively.
Immediately above the characteristic cutoff the series gives zero contribution,
whereas the FVM leaves a nonzero numerical tail.  This is the parameter space in
which the analytical series is materially better than the FVM for spectral-shape
studies.

The one-zone solution used in Fig.~\ref{fig:solution_comparison} can be written
as
\begin{align}
  N_{\rm int}(E,r,T)=
  {}&\frac{b(E_s)}{b(E)}Q(E_s)
  \frac{1}{2\pi^2r}
  \nonumber\\
  &\times
  \int_0^\infty dk\,k\sin(kr)\,
  e^{-k^2\Lambda(E,E_s)} .
  \label{eq:integral_solution}
\end{align}
Analytically this is equivalent to the positive closed form
\begin{equation}
  N_{\rm int}(E,r,T)=
  \frac{b(E_s)}{b(E)}
  Q(E_s)
  \frac{\exp[-r^2/(4\Lambda)]}{(4\pi\Lambda)^{3/2}} .
  \label{eq:integral_closed}
\end{equation}
For comparison with the two-zone series we also evaluate the two-zone integral
kernel of Osipov et al.~\cite{Osipov2020}, as used for pulsar positron
transport in, e.g., Ref.~\cite{Schroer2023}.  For an observer outside the
inhibited region, $r\geq R_1$, it is
\begin{align}
  N_{\rm Osi}(E,r,T)
  ={}&
  \frac{b(E_s)}{b(E)}Q(E_s)
  \frac{\xi}{\pi^2\Lambda_1 r}
  \nonumber\\
  &\times
  \int_0^\infty d\psi\,
  \frac{e^{-\psi}{\cal D}_2(\psi)}
  {A^2(\psi)+B^2(\psi)},
  \label{eq:osipov_kernel}
\end{align}
where $\xi=\sqrt{D_1/D_2}$, $\chi=\sqrt{\psi}\,R_1/\sqrt{\Lambda_1}$, and
\begin{align}
  {\cal D}_2(\psi)
  ={}&
  A(\psi)\sin\!\left(\sqrt{\psi}\,\frac{r\xi}{\sqrt{\Lambda_1}}\right)
  +B(\psi)\cos\!\left(\sqrt{\psi}\,\frac{r\xi}{\sqrt{\Lambda_1}}\right),
  \nonumber\\
  A(\psi)
  ={}&
  \xi\cos\chi\cos(\xi\chi)
  +\sin\chi\sin(\xi\chi)
  \nonumber\\
  &+
  \frac{1-\xi^2}{\xi\chi}\sin\chi\cos(\xi\chi),
  \nonumber\\
  B(\psi)
  ={}&
  \frac{\sin\chi-A(\psi)\sin(\xi\chi)}{\cos(\xi\chi)} .
  \label{eq:osipov_ab}
\end{align}
Here $\Lambda_1=\int_E^{E_s}D_1(E')dE'/b(E')$.  This integral solution assumes
an infinite outer region, whereas the series solution enforces the finite
$R_2$ boundary exactly.

\section{Population geometry and collective contribution}
\label{sec:geometry}

For a single impulsive source, the flux at Earth is proportional to
$N(E,r,T)$, where $T$ is both the source age and the diffusion time since the
instantaneous release of pairs.  For a population of identical burst-like
sources with locally uniform three-dimensional space density and constant birth
rate, the contribution from a radial shell is instead
\begin{equation}
  d{\cal F}_{\rm 3D}(E)\propto
  4\pi r^2 dr
  \int_{T_{\min}}^{T_{\max}} dT\,N(E,r,T).
  \label{eq:shell3d}
\end{equation}
The age integral is a continuous integral over this diffusion time, with a
uniform birth-rate weight between $T_{\min}$ and $T_{\max}$; it is not a
continuous spin-down injection calculation.  The factor $4\pi r^2dr$ is the
basic reason distant sources can matter.  In the figures we also show the
contribution per logarithmic distance interval,
$d{\cal F}/d\ln r\propto4\pi r^3\int dT\,N$.

The three-dimensional uniform case is not the only relevant geometry.  Galactic
pulsars are distributed in a disk, whose local source count grows only linearly
with projected distance.  In the thin-disk limit the contribution from an
annulus centered on the observer is
\begin{equation}
  d{\cal F}_{\rm disk}(E)\propto
  2\pi R\,dR
  \int_{T_{\min}}^{T_{\max}} dT\,N(E,R,T).
  \label{eq:thindisk}
\end{equation}
For a disk with vertical probability density $p(z)$, this becomes
\begin{align}
  d{\cal F}_{\rm disk}(E)
  \propto{}&
  2\pi R\,dR
  \int dT\int dz\,p(z)
  \nonumber\\
  &\times
  N\!\left(E,\sqrt{R^2+z^2},T\right).
  \label{eq:thickdisk}
\end{align}
We use an exponential vertical profile
$p(z)\propto\exp(-|z|/h)$ and show results for $h=0.2\,{\rm kpc}$.  The
thin-disk result is useful as a limiting case, but it overemphasizes very small
projected radii in a continuous-source approximation.  A real pulsar population
is discrete and has a finite nearest-neighbor distance, so the continuous disk
limit should not be interpreted literally below a few hundred parsecs.
For this reason, we quote the finite-thickness disk as the more physical disk
benchmark and use the thin disk only to understand the limiting geometry.

Our fiducial parameters are listed in Table~\ref{tab:parameters}.  We fix
$D_2(E)=3\times10^{28}(E/{\rm GeV})^{1/3}\,{\rm cm^2\,s^{-1}}$ and choose
$D_1(100\,{\rm TeV})=4.5\times10^{27}\,{\rm cm^2\,s^{-1}}$, which gives
$\eta=309.4$ for the common Kolmogorov energy dependence.  The inner-zone radius
is $R_1=50\,{\rm pc}$ and the outer numerical boundary is set to
$R_2=20\,{\rm kpc}$.  The injection is impulsive, with
$Q(E)\propto(E/10\,{\rm GeV})^{-1.9}\exp(-E/100\,{\rm TeV})$.  The source ages,
equivalently the diffusion times after the impulsive release, are integrated
uniformly from $0.03$ to $10\,{\rm Myr}$.

\begin{table*}[t]
\caption{Fiducial parameters for the shell-volume diagnostic.}
\label{tab:parameters}
\begin{ruledtabular}
\begin{tabular}{lc}
Parameter & Value \\
\hline
$R_1$ & $0.05\,{\rm kpc}$ \\
$R_2$ & $20\,{\rm kpc}$ \\
$D_2(E)$ & $3\times10^{28}(E/{\rm GeV})^{1/3}\,{\rm cm^2\,s^{-1}}$ \\
$D_1(100\,{\rm TeV})$ & $4.5\times10^{27}\,{\rm cm^2\,s^{-1}}$ \\
$\eta=D_2/D_1$ & $309.4$ \\
$B$ & $3\,\mu{\rm G}$ \\
ISRF & CMB $+$ IR $+$ optical blackbodies \\
$\gamma$ & $1.9$ \\
$E_{\rm cut}$ & $100\,{\rm TeV}$ \\
age range & $0.03$--$10\,{\rm Myr}$
\end{tabular}
\end{ruledtabular}
\end{table*}

The three-dimensional result is summarized in Table~\ref{tab:shell}.  The median
contributing radius is $r_{50}\simeq1.9\,{\rm kpc}$ at $10\,{\rm GeV}$,
$2.4\,{\rm kpc}$ at $30\,{\rm GeV}$, and $1.8\,{\rm kpc}$ at
$100\,{\rm GeV}$.  Less than one quarter of the $10$--$100\,{\rm GeV}$
contribution comes from within $1\,{\rm kpc}$.  At $300\,{\rm GeV}$ the
distribution begins to contract, with $r_{50}=1.23\,{\rm kpc}$.  At
$1\,{\rm TeV}$ the contribution is local: $65\%$ is inside $1\,{\rm kpc}$ and
$94\%$ is inside $2\,{\rm kpc}$.

\begin{table*}[t]
\caption{Age-integrated shell contribution for a three-dimensional uniform
population of impulsive pulsar sources.
Here $r_{50}$, $r_{80}$, and $r_{90}$ enclose 50\%, 80\%, and 90\% of the
radial integral in Eq.~\eqref{eq:shell3d}.}
\label{tab:shell}
\begin{ruledtabular}
\begin{tabular}{cccccc}
$E$ [GeV] & $r_{50}$ & $r_{80}$ & $r_{90}$ & $f(<1\,{\rm kpc})$ & $f(<2\,{\rm kpc})$ \\
\hline
10   & 1.91 & 3.14 & 3.86 & 0.22 & 0.53 \\
30   & 2.44 & 4.04 & 4.99 & 0.16 & 0.39 \\
100  & 1.85 & 3.12 & 3.88 & 0.24 & 0.54 \\
300  & 1.23 & 2.15 & 2.69 & 0.40 & 0.76 \\
1000 & 0.71 & 1.36 & 1.74 & 0.65 & 0.94
\end{tabular}
\end{ruledtabular}
\end{table*}

For a disk-like population the conclusion is softened but not erased.  In the
finite-thickness disk calculation, with
$h=0.2\,{\rm kpc}$, the median projected radius is $R_{50}=0.67\,{\rm kpc}$ at
$10\,{\rm GeV}$, $0.90\,{\rm kpc}$ at $30\,{\rm GeV}$, and $0.61\,{\rm kpc}$ at
$100\,{\rm GeV}$.  The fraction inside $1\,{\rm kpc}$ is therefore larger than
in the three-dimensional calculation, but it is not unity: $37$--$47\%$ of the
$10$--$100\,{\rm GeV}$ flux still comes from beyond $1\,{\rm kpc}$ in this
fiducial disk model.  At $1\,{\rm TeV}$ the disk result is strongly local, with
$89\%$ inside $1\,{\rm kpc}$.

\begin{acknowledgments}
During the preparation of this work, the authors used GLM5.2 to improve
readability and language. After using this tool, the authors reviewed and
edited the content as needed and take full responsibility for the content of
the publication. This work is supported by by the National Science Foundation of China (NSFC) through grant No. 12393853 and K. C. Wong Educational
Foundation. Y. Bao thanks Z. Liu for coding the proton version of the solution, and R. Liu and Benedikt Schroer for helpful discussions.
\end{acknowledgments}

\bibliographystyle{apsrev}
\bibliography{TZE}

\end{document}